\documentclass[shortnote,twocolumn]{jpsj3}
\usepackage{txfonts}
\usepackage{color}
\usepackage{ulem}

\title{
Enhanced Superconductivity in Close Proximity to the Structural Phase Transition of Sr$_{1-x}$Ba$_{x}$Ni$_{2}$P$_{2}$
}

\author{
Kazutaka Kudo\thanks{E-mail: kudo@science.okayama-u.ac.jp}, 
Yutaka Kitahama, 
Keita Iba, \\
Masaya Takasuga, 
and Minoru Nohara\thanks{E-mail: nohara@science.okayama-u.ac.jp}
}

\inst{
Research Institute for Interdisciplinary Science,\\ Okayama University, Okayama 700-8530, Japan
}

\abst{
The structural evolution and superconductivity of a 122-type solid solution Sr$_{1-x}$Ba$_{x}$Ni$_{2}$P$_{2}$ were studied. 
We found that an orthorhombic-tetragonal structural phase transition takes place at $x$ = 0.5, and is characterized by the P-P dimers breaking.  
The superconducting transition temperature exhibited its highest value of 2.85 K at $x$ = 0.4. 
}

\begin{document}
\maketitle

The transition-metal compounds AT$_2$X$_2$ exhibit a variety of crystal structures, and the most common is the tetragonal ThCr$_2$Si$_2$-type (space group $I4/mmm$, $D_{4h}^{17}$, No. 139). \cite{Hoffmann}
This structure can further be classified into two types: 
the ``collapsed" tetragonal that is characterized by the formation of molecule-like X-X dimers along the crystallographic $c$-axis, and the ``uncollapsed" tetragonal in which the X-X dimers are broken. 
Interestingly, some compounds exhibit structural phase transitions from the uncollapsed to collapsed tetragonal phases when hydrostatic pressure or chemical doping is applied. 
Typical examples include CaFe$_2$As$_2$ under pressure, \cite{Kreyssig,Goldman} CaFe$_2$(As$_{1-x}$P$_{x}$)$_2$, \cite{Kasahara} and Ca(Fe$_{1-x}$Rh$_{x}$)$_2$As$_2$, \cite{Danura} in which the formation of X-X bonds results in the loss of the Fe magnetic moment and the disappearance of superconductivity. 
Other notable examples are Sr$_{1-x}$Ca$_{x}$Co$_2$P$_2$,\cite{Jia} LaCo$_2$(Ge$_{1-x}$P$_{x}$)$_2$,\cite{Jia1} and SrCo$_2$(Ge$_{1-x}$P$_{x}$)$_2$,\cite{Jia2} in which the ferromagnetic quantum critical point is induced by the X-X dimer breaking. \cite{Jia2}

A rare but intriguing structure among AT$_2$X$_2$ appears in SrNi$_2$P$_2$. 
This compound crystallizes in an orthorhombic structure with the space group $Immm$ ($D_{2h}^{25}$, No. 71) \cite{Keimes}. 
In this structure, the unit cell is made up of 12 phosphorus atoms, of which four phosphorus atoms form dimers while the remaining eight do not. This results in a superstructure along the crystallographic $b$-axis, and hence causes orthorhombic distortion, as shown in Fig. 1(a). 
On the other hand, BaNi$_2$P$_2$ crystallizes in an uncollapsed tetragonal structure ($I4/mmm$), \cite{Keimes} in which no P-P dimers are formed, as shown in Fig. 1(b). 
Both compounds exhibit superconductivity, and the superconducting transition temperatures $T_{\rm c}$ are 1.4 K in SrNi$_2$P$_2$\cite{Ronning} and 2.5 K in BaNi$_2$P$_2$\cite{Mine,Tomioka}. 
This raises questions regarding which composition causes the P-P dimer breaking to occur in the solid solution Sr$_{1-x}$Ba$_{x}$Ni$_2$P$_2$, and how $T_{\rm c}$ varies as a function of the composition $x$.

In this paper, we present the results of our synthesis of the SrNi$_2$P$_2$--BaNi$_2$P$_2$ solid solution along with the crystallographic and superconducting properties of the samples. 
Our observations suggest that there was interplay between the structural transitions and superconductivity in Sr$_{1-x}$Ba$_{x}$Ni$_2$P$_2$. 

\begin{figure}[t]
\begin{center}
\includegraphics[width=8cm]{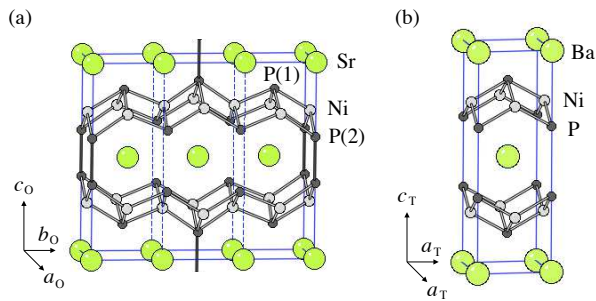}
\caption{
(Color online) The crystal structure of (a) SrNi$_2$P$_2$ (space group $Immm$, $D_{2h}^{25}$, No. 71) and (b) BaNi$_2$P$_2$ ($I4/mmm$, $D_{4h}^{17}$, No. 139). 
}
\end{center}
\end{figure}

\begin{figure}[t]
\begin{center}
\includegraphics[width=7cm]{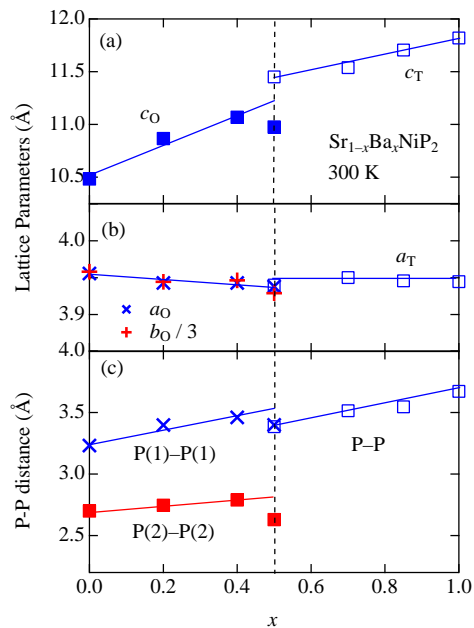}
\caption{
(Color online) The $x$ dependencies of the lattice parameters and the interlayer P-P distances in Sr$_{1-x}$Ba$_x$Ni$_2$P$_2$ at room temperature. 
The subscripts T and O denote the tetragonal and orthorhombic phases, respectively. See Fig. 1 for P(1) and P(2) of the orthorhombic phase. 
}
\end{center}
\end{figure}

Polycrystalline samples of Sr$_{1-x}$Ba$_x$Ni$_2$P$_2$ with nominal Ba content of 0.00 $\le$ $x$ $\le$ 1.00 were synthesized by means of a solid-state reaction. The stoichiometric amounts of the starting materials Ba, Sr, Ni, and P were mixed, ground, placed in an alumina crucible, and then sealed in an evacuated quartz tube. The ampoule was heated at 300 $^\circ$C for 5 h, at 700 $^\circ$C for 3 h, and at 1000 $^\circ$C for 24 h, followed by natural cooling in the furnace. Powder X-ray diffraction (XRD) studies confirmed that the resulting samples were a single phase of Sr$_{1-x}$Ba$_x$Ni$_2$P$_2$. The lattice parameters and interlayer P-P distances were estimated by the Rietveld refinement using the RIETAN-FP program. \cite{Izumi} The magnetization $M$ was measured using the Magnetic Property Measurement System (MPMS) by Quantum Design at temperatures above 1.8 K.

\begin{figure}[t]
\begin{center}
\includegraphics[width=8.5cm]{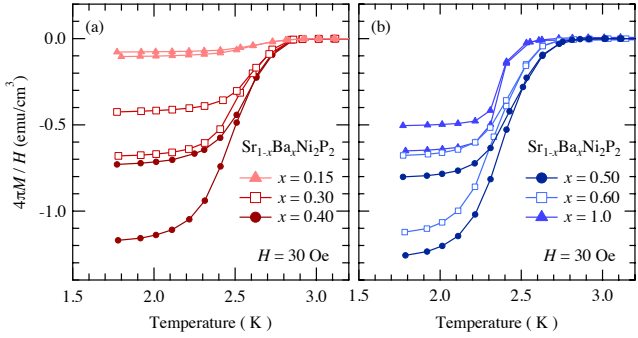}
\caption{
(Color online) The temperature dependence of the magnetization $M$ divided by the magnetic field $H$ for Sr$_{1-x}$Ba$_x$Ni$_2$P$_2$ with (a) $x$ $<$ 0.50 and (b) $x$ $\ge$ 0.50 at $H=$ 30 Oe in the zero-field and field cooling conditions. No correction for the diamagnetizing field was made.
}
\end{center}
\end{figure}

\begin{figure}[t]
\begin{center}
\includegraphics[width=7cm]{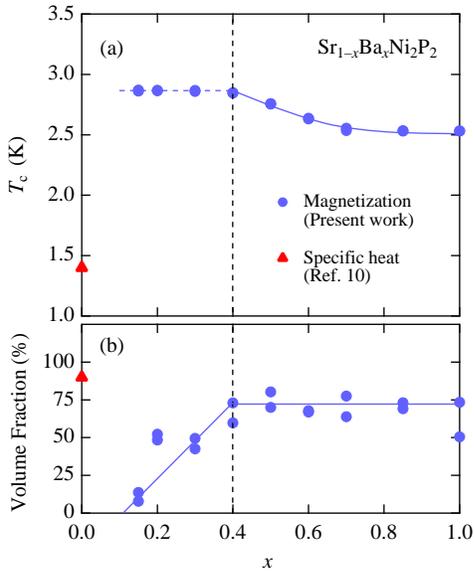}
\caption{
(Color online) The $x$ dependencies of (a) the superconducting transition temperature $T_{c}$ and (b) the superconducting volume fraction determined from the magnetization data measured in the field cooling conditions in Sr$_{1-x}$Ba$_x$Ni$_2$P$_2$ (blue circles). The $T_{\rm c}$ and volume fraction for the end member SrNi$_2$P$_2$ (red triangles) were determined from specific heat data in Ref. \citen{Ronning}. 
}
\end{center}
\end{figure}
%

Based on the XRD data measured at room temperature, the structure at $x <$ 0.5 was confirmed to be the orthorhombic structure ($Immm$), whereas at $x >$ 0.5 the structure was confirmed to be the uncollapsed tetragonal ($I4/mmm$). 
The orthorhombic--tetragonal structural transition can be seen in the $x$ dependence of the $c$ parameter, which exhibited a jump at $x$ = 0.5, as shown in Fig.~2(a).  
On the other hand, the variation in the $a$ parameter was very small, as shown in Fig.~2(b). 
The orthorhombic phase at $x <$ 0.5 can be characterized by the two distinct P-P distances, P(1)-P(1) and P(2)-P(2), as shown in Fig.~2(c). 
The P(2)-P(2) distance was comparable to the P-P distances of 2.251--2.710 {\AA} in the collapsed tetragonal phase of CaTM$_2$P$_2$ (TM = Fe, Co, Ni, Cu),  \cite{Hoffmann} and thus indicates a covalent bond. 
A large P(1)-P(1) distance suggests the absence of a covalent bond between them. 
In contrast, in the tetragonal phase at $x >$ 0.5, the P-P distance was uniform and much longer than that of the covalent P-P bond, as shown in Fig.~2(c), and the covalent P-P bonds were completely broken.

Figure 3 shows the temperature dependence of the magnetization $M$ for Sr$_{1-x}$Ba$_x$Ni$_2$P$_2$ at a magnetic field of 30 Oe in the zero-field and field cooling conditions. 
Using this data, we estimated the superconducting transition temperature $T_{\rm c}$ and superconducting volume fraction as a function of the doping $x$, as summarized in Fig.~4. 
Bulk superconductivity with a substantial superconducting volume fraction was observed at $x$ $\geq$ 0.4 for Sr$_{1-x}$Ba$_x$Ni$_2$P$_2$.
The superconducting transition temperature exhibited the highest value of $T_{\rm c}$ = 2.85 K at $x$ = 0.4. 
This $T_{\rm c}$ value was twice as high as the $T_{\rm c}$ = 1.4 K of the end member SrNi$_2$P$_2$. \cite{Ronning}
$T_{\rm c}$ decreased monotonically with increasing $x$, and 
the end member BaNi$_2$P$_2$ exhibited $T_{\rm c}$ = 2.53 K, which is in good agreement with previous reports. \cite{Mine,Tomioka}
In contrast, the superconductivity at 0.15 $\leq$ $x$ $<$ 0.4 is not bulk in nature.
The superconducting volume fraction decreased with decreasing $x$ below $x$ = 0.4, and the $T_{\rm c}$ was almost independent of $x$. 
This behavior suggests the coexistence of superconducting and non-superconducting phases, and their phase boundary is located at $x$ $\simeq$ 0.4. 
However, the XRD data suggested the formation of a continuous solid solution in this doping range. The orthorhombic--tetragonal structural phase boundary was located at $x$ $\simeq$ 0.5, as shown in Fig. 2. 
To account for this discrepancy, another phase transition, presumably to the collapsed tetragonal structure, should be invoked below room temperature.  
Incidentally, SrNi$_{2}$(P$_{1-x}$As$_{x}$)$_{2}$ is known to exhibit three structural phases, namely the orthorhombic ($Immm$), collapsed, and uncollapsed tetragonal ($I4/mmm$) structures, depending on the temperature and composition. \cite{Keimes2}

In conclusion, we have demonstrated that the SrNi$_2$P$_2$-BaNi$_2$P$_2$ solid solution exhibits a structural phase transition that is characterized by the breaking of the P(2)-P(2) dimers of the orthorhombic structure. 
The superconducting transition temperature was enhanced up to $T_{\rm c}$ = 2.85 K when the composition was in the vicinity of the structural transition. 
This $T_{\rm c}$ was higher than those of the end members SrNi$_2$P$_2$ (1.4 K) and BaNi$_2$P$_2$ (2.5 K). 
Thus, the tuning of the X-X bonding state in AT$_2$X$_2$ with the ThCr$_2$Si$_2$-type structure is an effective means to develop superconductors with enhanced $T_{\rm c}$.

\acknowledgment
A part of this work was performed at the Advanced Science Research Center, Okayama University.
This work was partially supported by Grants-in-Aid for Scientific Research (No. 26287082, 15H01047, 15H05886, and 16K05451) provided by the Japan Society for the Promotion of Science (JSPS) 
and the Program for Advancing Strategic International Networks to Accelerate the Circulation of Talented Researchers from JSPS.

\end{document}